\documentclass[preprint,aps,a4paper, showpacs]{revtex4}
\usepackage{graphicx}

\begin{document}
\draft
\title{Epidemics in small world networks}
\author{M. M. Telo da Gama and A. Nunes}
\address{Centro de F{\'\i}sica Te{\'o}rica e Computacional and 
Departamento de F{\'\i}sica, Faculdade de Ci{\^e}ncias da Universidade de 
Lisboa, P-1649-003 Lisboa Codex, Portugal}

\begin{abstract}

For many infectious diseases, a small-world network on an underlying regular
lattice is a suitable simplified model for the contact structure of the host
population. It is well known that the contact network, described in this 
setting by a single parameter, the small-world parameter $p$, plays an 
important role both in the short term and in the long term dynamics of 
epidemic spread.
We have studied the effect of the network structure on models of immune 
for life diseases and found that in addition to the reduction of the 
effective transmission rate, through the screening of infectives, spatial 
correlations may strongly enhance the stochastic fluctuations. As a 
consequence, time series of unforced Susceptible-Exposed-Infected-Recovered 
(SEIR) models provide patterns of recurrent epidemics with realistic 
amplitudes, suggesting that these models together with complex networks of 
contacts are the key ingredients  to describe the prevaccination dynamical 
patterns of diseases such as measles and pertussis. 
We have also studied the role of the host contact strucuture in 
pathogen antigenic variation, through its effect on the 
final outcome of an invasion by a viral strain of a population where a 
very similar virus is endemic. Similar viral strains are modelled 
by the same infection and reinfection parameters, and by a given degree of 
cross immunity that represents the antigenic distance between the 
competing strains. We have found, somewhat surprisingly, that clustering 
on the network decreases the potential to sustain pathogen diversity. 

\end{abstract}


\maketitle

\section{introduction}

Nearly a decade ago, Watts and Strogatz introduced a class of networks
with a topology interpolating between that of lattices and random graphs
\cite{w&s}. In these models a fraction of the links of the lattice
is randomized by connecting nodes, with probability $p$, with any other
node. For a range of $p$ the network exhibits 'small world' behaviour, 
where a local neighbourhood (as in lattices) coexists with a short average 
path length (as in random graphs). Analysis of real networks \cite{realnets} 
reveals the existence of small worlds in many interaction networks, including 
networks of social contacts. Recently, attention has been focussed on the 
impact of network topology on the dynamics of the processes running on it 
with emphasis on the spreading of infectious diseases.
Early numerical and analytical studies have focussed on the calculation of 
epidemic thresholds
\cite{moore&newman,thresholds} for different classes of networks and on
the short term dynamics of epidemic bursts \cite{bursts} for small world
networks of the Watts and Strogatz type. By contrast, the effects of
network topology on the long term dynamics of epidemic spread has received
much less attention.

In the last couple of years we have studied simple epidemic models on 
dynamic (or annealed) Watts and Strogatz small world networks \cite{cftc} 
and found that, for
a large range of $p$, epidemics persist for times that approach those of
homogeneously mixed populations providing suitable models to investigate the
effects of spatial correlations on the long term dynamics of epidemic
spread. In our model the $N$ nodes of the network represent individuals. 
The links are then connections along which the infection spreads. If all 
the links are local (lattice) the persistence of the epidemic is low and the 
endemic state may not be reached \cite{bursts}. On the other hand, if 
the network is random (homogeneously mixed population), the persistence is 
high but the effect of spatial correlations is neglected \cite{stochastic}. 
The dynamic version of the Watts and Strogatz model \cite{cftc} is thus the 
simplest model of a real network of social contacts \cite{realnets}. 

Using a SEIR (Susceptible-Exposed-Infective-Recovered) model, we have obtained 
a quantitative description of the incidence oscillations that characterize 
the prevaccination recurrent epidemics of immune for life diseases such 
as measles and pertussis. The model, on the same network, was found to 
describe the qualitative differences observed in the long term dynamics of 
such diseases. 
We have also used a simple two-strain version of a recently proposed 
SIR model with reinfection to determine the outcome of an invasion 
in multiple strain infections. The results uncovered a surprising effect 
of the network structure on the diversity of pathogens, that was found to 
decrease, rather than increase, as the spatial correlations on the dynamic 
small world network increase. 

\section{recurrent epidemics on networks}

The Susceptible-Infected-Recovered (SIR) mean field model in a closed 
population of $N$ individuals is given by the equations 
\begin{eqnarray}
\dot S & = & -  \beta S I /N \nonumber \\
\dot I & = & \beta S I /N - \gamma I
\label{sir}
\end{eqnarray}
where $S$ (respectively $I$, $R$) is the number of susceptible 
(respectively infected, recovered) individuals and $\beta $
is the transmissibility. This is a model for the dynamics of a 
disease that confers permanent immunity and it is easily generalized to an 
epidemic that takes place on a network. As for models of epidemic spread on 
regular lattices \cite{Grassberger}, the model may be mapped onto percolation 
on the same network \cite{moore&newman}. The percolation transition corresponds 
to the epidemic threshold, above which an epidemic outbreak is possible (i.e. 
one that infects a non-zero fraction of the population, in the limit of large 
populations) and the size of the percolating cluster above this transition 
corresponds to the size of the epidemic.
The SEIR model is a generalisation of (\ref{sir}) that takes into account a 
latency period, during which individuals carry the pathogen but are not yet 
infective. This was found to be relevant in the modelling of childhood 
diseases \cite{am&murray}.  
 
The endemic persistence of a disease on a closed population requires 
renewal of susceptibles through demography or loss of immunity.
To assess the effect of network topology on the long term dynamics of 
simple spatial models for childhood diseases we implemented discrete SIR 
and SEIR models with births and deaths corresponding to an average life 
time of 61 years \cite{SIR,SEIR,cftc} on a square lattice with $N$ sites 
and dynamic small world interaction rules. 
We found characteristic medium and long-term dynamics related, in a 
quantitative fashion, to the structure of the network of contacts. In 
particular, as the small world parameter, $p$, decreases, 
the increase in spatial correlations (i) decreases the effective 
transmissibility through 
the screening of infectives and susceptibles, which in turn increases the 
value of the transmissibility at the endemic threshold. 
In addition, the spatial correlations (ii) enhance the stochastic 
fluctuations with 
respect to the homogeneously mixed stochastic model. This effect is 
particularly strong at 
low $p$, where the relative fluctuations are largest and where as a 
consequence (iii) the dependence of the steady state densities on the 
effective transmissibility predicted 
by the mean-field equations breaks down \cite{cftc}.


In Figure 1(a) we have plotted the incidence time series of a SEIR 
model \cite{SEIR} with the epidemiological parameters (latency period 
$\tau_l$ and recovery time $\tau_i$) taken for measles and two different  
values of the small world parameter $p$, $p=1$ and $p=0.2$.
For homogeneously mixed populations, $p=1$, the infective time series 
exhibits incidence oscillations with an average period 
of two years. However, the amplitude of these oscillations is underestimated 
when compared with the incidence oscillations of measles in real data 
for similar population sizes \cite{data}.
By contrast, in the infective time series for SEIR simulations on the 
network for $p=0.2$ the amplitude of the incidence peaks is shown to 
increase significantly, in line with the real data.

As with any unforced model, the characteristic biennial cycles 
of prevaccination records cannot be obtained without fine tuning of the 
model's parameters, transmissibility, $\beta$, and probability of 
long-range infection, $p$. 
Despite this fact, which means that, for measles, seasonality cannot be 
ignored \cite{grenfell}, the reported effects 
of spatial correlations should also be taken into account 
as a key issue in explaining incidence amplitudes.


Figure 1(b) shows the incidence time series of model \cite{SEIR}
with the epidemiological parameters taken for 
pertussis and three different values of the small world parameter $p$, 
$p=1$, $p=0.2$ and $p=0.1$. 
Again, we see that for the smaller value of $p$ the amplitude of the 
incidence peaks is significantly enhanced with respect to the 
homogeneously 
mixed case $p=1$. However, for intermediate values of $p$, and in 
particular for $p=0.2$, this effect is almost negligible.  

The main conclusion of the previous analysis is that the impact of the 
spatial correlations depends also on the stability of a particular disease, 
measured by the distance of the endemic phase to the corresponding threshold 
on the network, that is disease dependent \cite{tudor}.

\section{Evolutionary dynamics of multiple strains}

We consider the invasion by a pathogen strain of a population where another
strain is endemic. A single viral strain is meant to represent a cluster of 
co-circulating antigenically close strains, and its uncoupled dynamics is 
described by a SIR model with transmissibility $\beta $ for first infection
and partial immunity against reinfection. The 
reinfection parameter $\sigma$ of the model is the ratio between disease 
transmissibilities for reinfection and for first infection. It measures 
the size of the cluster, in terms of antigenic distance.

For realistic values of the birth rate, the SIR model with 
partial immunity  exhibits an abrupt
change in the levels of infection as $\beta $ increases across a
transmissibility threshold. This region of 
abrupt change of infection prevalence  
is related to a bifurcation that occurs in the
model with zero birth rate {\cite{cftc}}, and we shall use
the term reinfection threshold for the smoothed transition 
that takes place in models with small birth rates.   
Recent studies {\cite{gg,others}} have revealed a connection between the 
reinfection threshold and the potential for 
pathogen diversity, that increases as the level of 
infection increases.




In two strain dynamics, disease transmissibility for infection
by one of the strains of individuals previously infected by the other strain
is also reduced by a factor $\sigma _{\times}$, due to cross-immunity. 
We assume that cross-immunity is weaker than within strain partial immunity, 
reflecting the fact that the antigenic distance between clusters must 
be larger than the cluster size.
Therefore, we take $\sigma < \sigma _{\times}$, and $\delta = \sigma 
_{\times} - \sigma $ is a measure of the difference between the two 
competing strains, in terms of the immune response that they trigger.

For a model with two (symmetrical) strains we have found that, at the 
level of mean field approximation, the reinfection threshold is the 
boundary between two different regimes. For similar strains 
($\delta \approx 0.02$), 
the behaviour of the model below threshold is strain replacement, and the 
behaviour on and above threshold is strain coexistence.
For distinct viral strains ($\delta \approx 0.2$), the outcome is always 
strain coexistence, but the density oscillations are negligible above 
threshold, and very pronounced below threshold, implying that taking 
into account that the population is discrete, for realistic population 
sizes we find extinction of both strains below threshold {\cite{cftc}}.
This scenario has been confirmed by simulations of a stochastic model for 
homogeneously mixed populations{\cite{Gokaydin}}. In the following 
we consider the contact structure of the host population and investigate its 
effect on pathogen diversity and evolution. 

We performed individual based simulations starting from an initial 
condition where the system is close to the steady state for a single 
resident strain and introduced a small fraction of individuals infected by 
the invading strain. We found that as local effects become important 
(low $p$), strain replacement (and also total 
extinction) are favoured with respect to the homogeneously mixed regime 
($p=1$). 
This happens both for similar ($\delta \approx 0.02$) and for distinct
strains ($\delta \approx 0.2$), and, as opposed to the effect reported
in the preceeding section, does not require $p$ to be close to the epidemic
persistence transition, only that the competing strains have moderate
transmissibilities (see Figure 2).
 



Our results differ from those reported in 
\cite{Buckee} for a model of strain evolution, with a high rate of 
supply of naive susceptibles, on a (static) small world network of 
contacts,  where spatial correlations were found to favour pathogen 
diversity by reducing the spread of acquired immunity.
In our model, with a low rate of supply of naive susceptibles, 
the main effect of spatial correlations is the screening of 
infectives that leads to a reduction of the effective transmissibility. As 
$p$ decreases, this effect brings a system above the reinfection threshold 
closer to or even below the reinfection threshold, also favouring replacement 
for similar strains and either replacement or extinction for distinct 
ones. 


Since screening of infectives occurs quite generally in structured 
populations, the effect reported here should be fairly independent of the 
network's detailed topology and correlations.

\section{Conclusions}

Our results for the SEIR model support the conclusion that, for some purposes, 
successful modeling of disease spread must take into 
account that populations are finite and discrete, and must include a realistic 
representation of the spatial degrees of freedom or, more generally, of the 
interaction network topology. 
We suggest a reassessment of the impact of environmental forcing 
by studying its effects on 
a realistic autonomous model as the one that 
we propose. In particular, fluctuation enhancement by spatial correlations 
might remove some of the constraints on the strength of seasonality and avoid 
fine tuning of seasonal forcing amplitudes.  

In the context of multiple strain infections and the evolution and 
diversity of pathogens, the results of our simulations strongly support 
the conclusion that in systems with reinfection and low rates 
of supply of naive susceptibles, the major effect of the host population 
structuring is a reduction of pathogen diversity.  

\section {Acknowledgements}
Financial support from the Portuguese Foundation for Science and 
Technology (FCT) under contract POCTI/ESP/44511/2002 is gratefully 
acknowledged. The authors also acknowledge the contributions of J. P. 
Torres 
and M. Sim\~oes to test and to improve the code used in the simulations.


\centerline{FIGURES}

{\bf FIGURE 1}
Infectives time series for homogeneously mixed and 
spatially structured populations from SEIR 
simulations on $N = 1000 \times 1000$ lattices for
a) measles ($\tau_i = 8$ days, $\tau_l = 6$ days) and
b) pertussis ($\tau_i = 18$ days, $\tau_l = 8$ days).
a) Grey line: Results for a homogeneously mixed population, $p=1$, and 
transmission rate $\beta= 2.4$ day $^{-1}$. 
Black line: Results for a network with $p=0.2$ and $\beta=4.75$ day$^{-1}$ 
also exhibiting an average period close to two years but much larger 
incidence oscillations. 
b) From top to bottom: Results for a homogeneously mixed population, $p=1$, 
and transmission rate $\beta= 1.5$ day $^{-1}$;
Results for a network with $p=0.2$ and $\beta=4.0$ 
day$^{-1}$; Results for a network with $p=0.1$ and $\beta=7.0$  
day$^{-1}$. 
As for measles, the transmissibility is tuned to preserve 
the average period of the homogeneously mixed population. The increase 
in the incidence peak's amplitude is negligible for $p=0.2$, but 
for $p=0.1$ the fluctuations 
are significantly enhanced, as for measles on 
a network with $p=0.2$.

\bigskip

{\bf FIGURE 2} Times series for invasion dynamics simulations of 
evolved viral strains, on small world networks over lattices with 
$N = 800 \times 800$ nodes. The vertical axis measures the logarithm of
the densities of infectives of the circulating and of the invading
strains. We have taken $1/80$ year$^{-1}$ for the population birth rate, 
$203$ year$^{-1}$ for the transmissibility, $0.27$ for the reinfection 
parameter 
and $52$ year$^{-1}$ for the recovery rate of both diseases. We have 
considered, in (a), similar strains, with $\delta =0.02$, and, in (b),
antigenically distant strains with $\delta =0.2$, and taken
$p=1$ (resp. $p=0.3$, $p=0.1$) in the first (resp. second, third)
column.
 
\newpage

\begin{figure}
\begin{center}
\begin{tabular}{cc}
\includegraphics{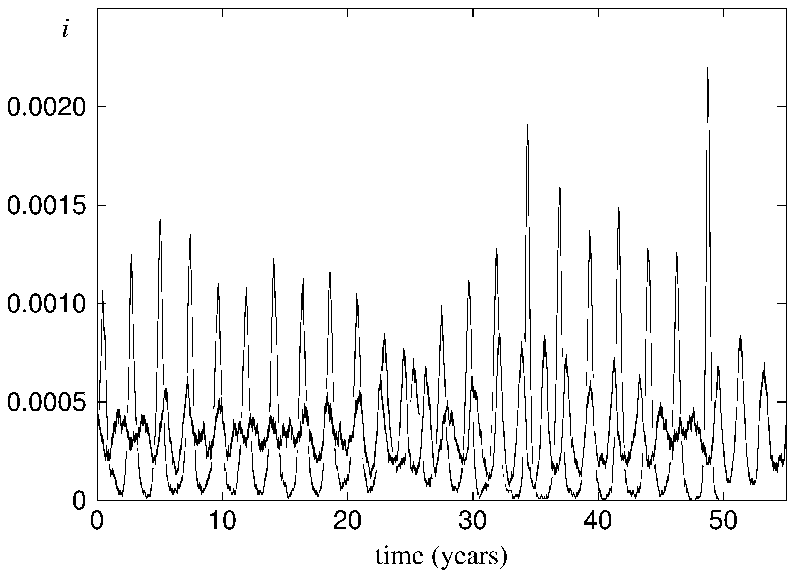}
&
\includegraphics{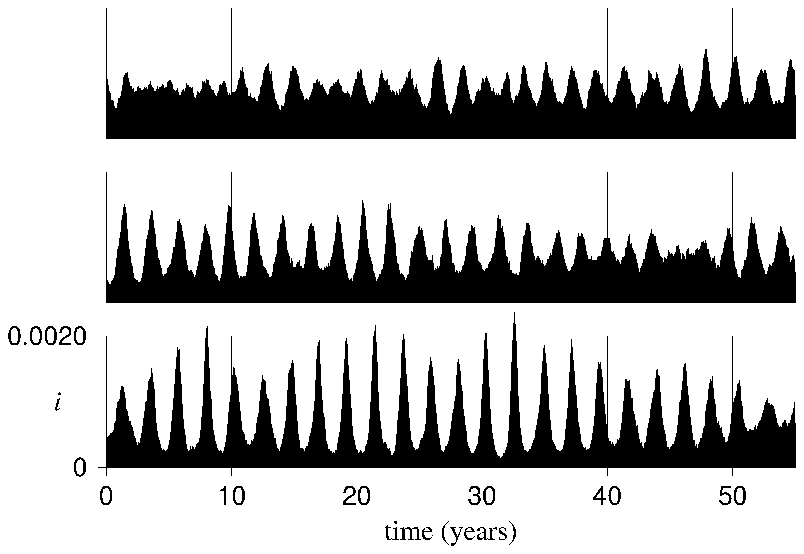}
\end{tabular}
\end{center}
\caption{}
\label{fig1}
\end{figure}



\begin{figure}
\begin{center}
\begin{tabular}{ccc}
\includegraphics[height=.25\textheight,angle=0]{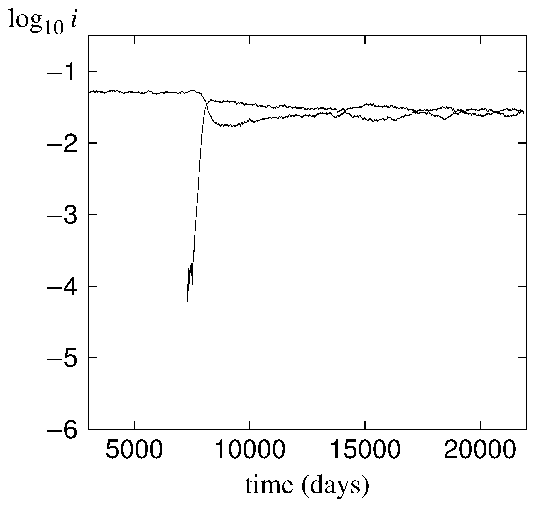}
&
\includegraphics[height=.25\textheight,angle=0]{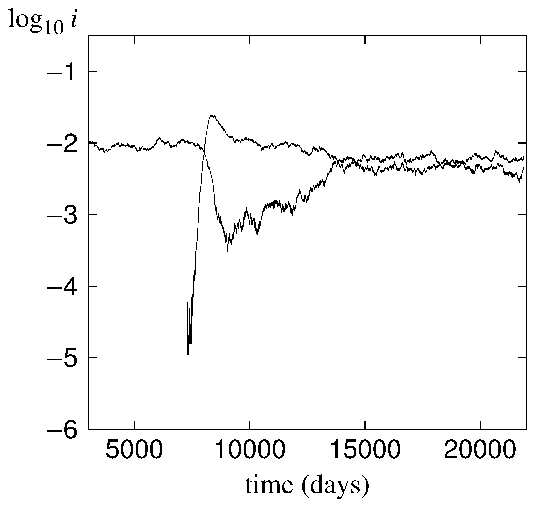}
&
\includegraphics[height=.25\textheight,angle=0]{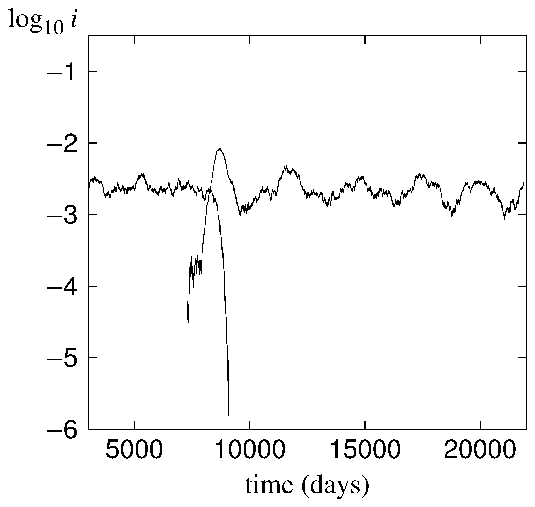}
\\
\includegraphics[height=.25\textheight,angle=0]{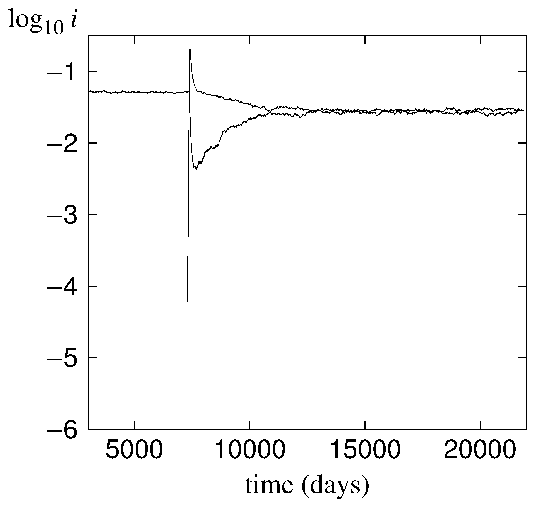}
&
\includegraphics[height=.25\textheight,angle=0]{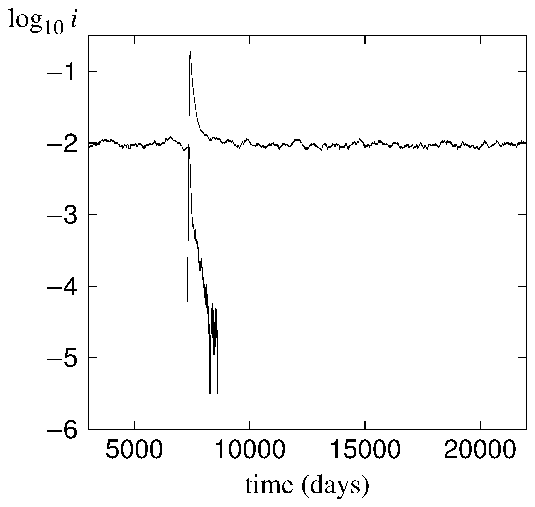}
&
\includegraphics[height=.25\textheight,angle=0]{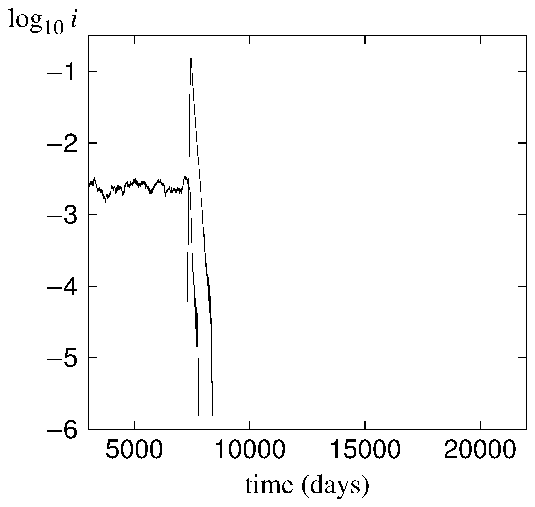}
\end{tabular}
\end{center}
\caption{}
\label{fig2}
\end{figure}
 

\begin{thebibliography}{99}
\bibitem{SIR}{A community of $N$ (fixed) individuals comprises, at time 
$t$, $S$ susceptibles, $I$ infectives in circulation and $R$ recovered or 
removed. Constant infection, $\beta$, and birth/death, $\mu$, rates are 
assumed. We consider a square lattice of size $N=L^{2}$ 
with periodic boundary conditions. The (random) variables at each site may 
take one of the values: $S$, $I$ or $R$. We account for local connections 
with $k$ neighbouring sites, with $k=12$, and long-range connections, with 
a small world probability, $p$. Birth, death and infection occur 
stochastically, with fixed rates ($\mu$, $\mu$, $\beta$) while 
recovery is deterministic after $\tau_i$ time steps. The recovery time 
sets the time scale.}
\end{thebibliography}
\end{document}